\documentstyle[floats,aps,epsf]{revtex}

\begin{document}
\draft

%%XXX \special{src: 18 MHJPISA3.TEX} %Inserted by TeXtelmExtel

%%XXX \special{src: 21 MHJPISA3.TEX} %Inserted by TeXtelmExtel

\newcommand{\lsix} {La$_{1.94}$Sr$_{0.06}$CuO$_4$}
\newcommand{\lsco} {La$_{2-x}$Sr$_x$CuO$_4$}
\newcommand{\la} {$^{139}$La}
\newcommand{\cu} {$^{63}$Cu}
\newcommand{\cuo} {CuO$_2$}
\newcommand{\ybco} {YBa$_{2}$Cu$_3$O$_{6+x}$}
\newcommand{\ybcoca} {Y$_{1-x}$Ca$_x$Ba$_2$Cu$_3$O$_6$}
\newcommand{\etal} {{\it et al.}}
\newcommand{\ie} {{\it i.e.}}

%%XXX \special{src: 33 MHJPISA3.TEX} %Inserted by TeXtelmExtel

\wideabs{

%%XXX \special{src: 37 MHJPISA3.TEX} %Inserted by TeXtelmExtel

%%XXX \special{src: 40 MHJPISA3.TEX} %Inserted by TeXtelmExtel

\title{\vspace*{-1cm}\hfill{\normalsize \tt 
\begin{flushleft}
\begin{tabular}{l}
\hline
Submitted to Applied Magnetic Resonance\\
(Proceedings of the Specialized Colloque Ampere\\
"EPR, NMR and NQR in Solid State Physics:\\
Recent Trends")\\
\hline
\end{tabular}
\end{flushleft}}
\vspace{1cm}
NQR Study of Spin-Freezing in Superconducting \lsco:\\ The Example of $x=0.06$}

%%XXX \special{src: 56 MHJPISA3.TEX} %Inserted by TeXtelmExtel

%%XXX \special{src: 59 MHJPISA3.TEX} %Inserted by TeXtelmExtel

\author{\sf M.-H. Julien$^{1,2,}$\protect\cite{mhj},
P. Carretta$^1$, F. Borsa$^{1,2}$}

%%XXX \special{src: 64 MHJPISA3.TEX} %Inserted by TeXtelmExtel

\address{\sf $^1$Dipartimento di Fisica "A. Volta", Unit\'a INFM di Pavia,
Via Bassi 6, I-27100 Pavia, Italy}
\address{\sf $^2$Department of Physics and Astronomy, Ames Laboratory,
Iowa State University, Ames IA-50011, USA}

%%XXX \special{src: 71 MHJPISA3.TEX} %Inserted by TeXtelmExtel

%%XXX \special{src: 74 MHJPISA3.TEX} %Inserted by TeXtelmExtel

\date{July 1999}
\maketitle

%%XXX \special{src: 79 MHJPISA3.TEX} %Inserted by TeXtelmExtel

\begin{abstract}

%%XXX \special{src: 83 MHJPISA3.TEX} %Inserted by TeXtelmExtel

{\sf
NQR \la~and \cu~spin-lattice relaxation rate (1/$T_1$) measurements in a
\lsix~single crystal are described.
Slowing-down of Cu$^{2+}$ spin fluctuations is evidenced 
through a dramatic increase of 1/$^{139}T_1$ on cooling.
While the onset of diamagnetism occurs at $T_c=8$~K,
1/$^{139}T_1$ has a peak at $T_g\simeq 5$~K, when the characteristic
frequency of magnetic fluctuations reaches the NQR frequency $\nu_Q\simeq19$~MHz.
In agreement with a number of previous studies,
these results show that the so-called "cluster spin-glass" phase persists
in the superconducting regime.
Issues concerning the coexistence of the two phases are discussed.
}

%%XXX \special{src: 99 MHJPISA3.TEX} %Inserted by TeXtelmExtel

\end{abstract}

%%XXX \special{src: 103 MHJPISA3.TEX} %Inserted by TeXtelmExtel

%\pacs{\sf PACS numbers: 76.60.-k, 74.25.Ha, 74.72.Dn}

%%XXX \special{src: 107 MHJPISA3.TEX} %Inserted by TeXtelmExtel

%%XXX \special{src: 110 MHJPISA3.TEX} %Inserted by TeXtelmExtel

}

%%XXX \special{src: 114 MHJPISA3.TEX} %Inserted by TeXtelmExtel

\narrowtext

%%XXX \special{src: 118 MHJPISA3.TEX} %Inserted by TeXtelmExtel

%%XXX \special{src: 121 MHJPISA3.TEX} %Inserted by TeXtelmExtel

\subsection{Introduction}
\label{introduction}

%%XXX \special{src: 126 MHJPISA3.TEX} %Inserted by TeXtelmExtel

Magnetic resonance techniques, 
nuclear quadrupole resonance (NQR), nuclear magnetic resonance
(NMR) and muon spin resonance ($\mu$SR), have been extensively used to establish
the phase diagram of high temperature cuprate superconductors.
Following early identifications of gross features \cite{Watanabe87,Kitaoka88,Budnick88},
these probes have rapidly revealed puzzling magnetic phenomena in
\lsco, at low hole doping \cite{Reviews}:

%%XXX \special{src: 136 MHJPISA3.TEX} %Inserted by TeXtelmExtel

- There is a magnetic freezing at low temperature {\it in the 3D ordered AF phase}
$0<x\leq 0.02$ \cite{Watanabe87,Chou93,Borsa95}.
All Cu$^{2+}$ moments become frozen below a freezing temperature $T_f\ll T_N$
which is proportional to the concentration $x$ of doped holes.

%%XXX \special{src: 143 MHJPISA3.TEX} %Inserted by TeXtelmExtel

- With increasing doping $0.02\leq x \leq 0.05$,
the N\'eel AF phase is replaced by a spin-glass phase at $T_g \propto \frac{1}{x}$
\cite{Harshman88,Sternlieb90,Chou95}.
However, it can be inferred from 
the muon precession signal below $T_g$ \cite{Budnick88,Harshman88,Niedermayer98} ,
from neutron scattering data \cite{Hayden91,Keimer92}
and from the \la~NQR spin-lattice relaxation rate ($1/T_1$) above $T_g$ \cite{Cho92},
that this phase still retains strong AF correlations limited by finite-size effects.
Thus, the frozen state was thought {\it not} to be a usual spin-glass \cite{Cho92},
but a "cluster spin-glass", \ie~frozen clusters of locally staggered 
magnetization, and random orientations of their quantization
axis  \cite{Emery93,Gooding97}.
Strong support to this picture was recently provided
by NMR measurements in \lsix~\cite{Julien99}.

%%XXX \special{src: 160 MHJPISA3.TEX} %Inserted by TeXtelmExtel

- Spin-freezing has been (rather often) observed in {\it superconducting} compounds
with typically $0.05<x\leq 0.10$
\cite{Kitazawa88,Weidinger89,Uemura89,Watanabe94,Oshugi94,Kukovitsky95}.
However, the claim of microscopically coexisting
superconducting and magnetic orders \cite{Kitazawa88,Weidinger89}
has been repeatedly criticized and attributed
to sample inhomogeneity \cite{Heffner89,Harshman89,Kiefl89}.
Recently, an important result was reported by Niedermayer \etal:
not only the persistence of the spin-glass phase in the
superconducting regime was confirmed in \lsco, but the same phase diagram
was also shown to hold for \ybcoca~\cite{Niedermayer98}
(see also \cite{Hodges91}).
Although inhomogeneity in the Ca or Sr concentration is unavoidable to some extent,
the results of Niedermayer \etal~\cite{Niedermayer98} indicate that
this is unlikely to explain the occurrence  of magnetic freezing far inside
the superconducting phase, in a similar way for both systems.

%%XXX \special{src: 179 MHJPISA3.TEX} %Inserted by TeXtelmExtel

These magnetic phases are not yet fully characterized and their
origin thus remains controversial.
Given the experimental evidence in favor of an unconventional
($d$-wave) superconductivity in the cuprates \cite{dwave},
and its plausible connection with AF correlations \cite{AFNMR,NMRreview,Dai99},
understanding these freezing phenomena appears obviously crucial.
These are relatively low temperature phases occurring at low doping,
but they appear in close contact to superconductivity in the phase diagram.

%%XXX \special{src: 190 MHJPISA3.TEX} %Inserted by TeXtelmExtel

%%XXX \special{src: 193 MHJPISA3.TEX} %Inserted by TeXtelmExtel

\subsection{Experiment}
\label{experiment}

%%XXX \special{src: 198 MHJPISA3.TEX} %Inserted by TeXtelmExtel

\lsix~is at the verge of the (underdoped)
superconducting phase, thus close to the cluster spin-glass phase.
Here, we describe NQR spin-lattice relaxation (1/$T_1$) measurements
of \la~and \cu~nuclei in an $x=0.06$
single crystal \cite{Lin97}.
Magnetization measurements have shown a superconducting transition
with an onset at $T_c=8$~K.
A comprehensive report of our NMR and NQR results 
in this compound can be found in Ref. \cite{Julien99}.
Here, we review the NQR relaxation results and discuss them in more details.

%%XXX \special{src: 211 MHJPISA3.TEX} %Inserted by TeXtelmExtel

%%%%%%%%%%%%%%%%%%%%%%%%%%%%%%

%%XXX \special{src: 215 MHJPISA3.TEX} %Inserted by TeXtelmExtel

\begin{figure*}[!t]
\vspace{-3cm}
\centerline{\epsfxsize=150mm \epsfbox{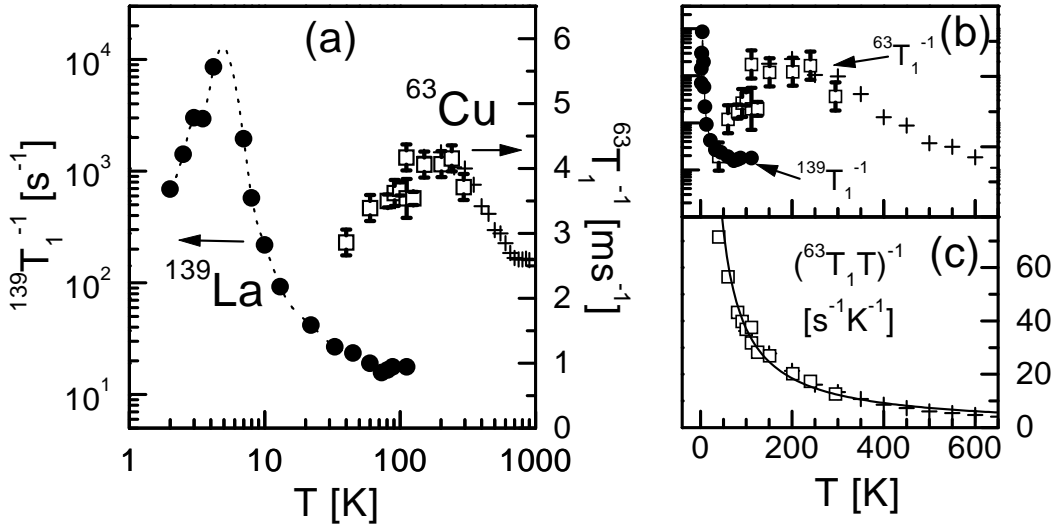}}
\vspace{-12cm}
\caption{
(a) $^{63}$Cu 1/$T_1$; this study (squares, right scale)
and from ref. \protect\cite{Fujiyama97} (crosses, right scale),
and $^{139}$La 1/$T_1$ (dots, left scale).
The peak in ($^{139}T_1)^{-1}$ defines the spin-freezing temperature $T_g\simeq 5$~K at
the NQR time scale $\sim 10^{-8}$ s.
(b) Same data as in (a), in the same vertical scale, but with a linear horizontal scale.
(c) $^{63}$Cu 1/$T_1$ data -symbols are the same as in (a,b).
The line is the function $f(T)=3700/T$.
}
\label{figure}
\end{figure*}
%%%%%%%%%%%%%%%%%%%%%%
When a nucleus of gyromagnetic ratio $\gamma$ is sensitive
to magnetic fluctuations through the hyperfine interaction $A(q)$,
1/$T_1$ can be expressed as:
%__________________
%
\begin{equation}
\frac{1}{T_1} = k_{\rm B} T\;\;\gamma^2
\sum_{q,\alpha\perp H} \frac{|A_\alpha(q)|^2}{(g_{\alpha}\mu_{\rm B})^2}
\;\;\frac{\chi^{\prime\prime}_\alpha(q,\omega_n)}{\omega_n}
\end{equation}
%_____________

%%XXX \special{src: 246 MHJPISA3.TEX} %Inserted by TeXtelmExtel

In most cases (in the absence of a gap), this can be rewritten as:
\begin{equation}
\frac{1}{T_1} = k_{\rm B} T\;\;\gamma^2
\sum_{q,\alpha\perp H} \frac{|A_\alpha(q)|^2}{(g_{\alpha}\mu_{\rm B})^2}
\;\;\frac{\chi^\prime_\alpha(q,0)}{\Gamma(q)}
\end{equation}
%%%
where $\Gamma(q)$ is the characteristic energy of spin fluctuations at
wavevector $q$.
When magnetic fluctuations slow down ($\Gamma(q)$
decreases) and provided their amplitude $\chi^\prime(q,\omega=0)$ is not much reduced
(in general, $\chi^\prime(q,\omega=0)$ rather increases) then $1/T_1$ increases.
This will be the main effect reported here.

%%XXX \special{src: 262 MHJPISA3.TEX} %Inserted by TeXtelmExtel

%%XXX \special{src: 265 MHJPISA3.TEX} %Inserted by TeXtelmExtel

\subsection{\cu~NQR results}
\label{curesults}

%%XXX \special{src: 270 MHJPISA3.TEX} %Inserted by TeXtelmExtel

The $^{63}$Cu nuclear spin-lattice relaxation rate 1/$^{63}T_1$ was measured
at the center of the NQR spectrum with full width at half maximum $\simeq 2$~MHz.
The recovery of the magnetization after a sequence
of saturating pulses could be fitted by a single exponential over at least
one decade at all temperatures.
The $T$-dependence of 1/$^{63}T_1$ is shown in Fig. 1(a,b),
along with data from Fujiyama \etal~\cite{Fujiyama97}.
Both sets of data agree very well in the
range $T$=100-300~K. As can be seen in the data of Fujiyama \etal,
1/$^{63}T_1$ increases with decreasing $T$ from 700~K to 250~K.

%%XXX \special{src: 283 MHJPISA3.TEX} %Inserted by TeXtelmExtel

This behaviour resembles other data at lower doping \cite{Imai93,Carretta99},
where the increase of $1/T_1$ with decreasing $T$ is ascribed to the increase 
of {\it two-dimensional} magnetic correlations. This regime is called the
"renormalized classical regime" \cite{Chakraverty89}.
As a matter of fact, spin correlations {\it in this intermediate $T$-range} evolve very
smoothly from the undoped insulator La$_2$CuO$_4$  to "metallic" \lsco~with $x$$\sim$0.06.
Actually, we have pointed out elsewhere \cite{Julien99} that a close
inspection of data in Ref. \cite{Fujiyama97} suggests that the
highest concentration at which the renormalized classical regime is observed is close to
$x=0.125=\frac{1}{8}$, where charge-stripes are best defined \cite{Hunt99}.
This is compatible with the idea of a magnetic quantum critical point at $x=\frac{1}{8}$,
as other experiments also suggest \cite{Hunt99,Aeppli97}.
Note that our observation relies on the work of Fujiyama \etal~
because it is the most comprehensive dataset at low doping reported to date \cite{Fujiyama97},
but there is no contradiction with other $^{63}T_1$ data in the literature
(in particular \cite{Imai93}).

%%XXX \special{src: 302 MHJPISA3.TEX} %Inserted by TeXtelmExtel

Below 250~K, 1/$^{63}T_1$ flattens and finally decreases below $\sim$150~K.
The $T$-dependence below 200~K could however not be accurately determined
since the Cu NMR signal becomes too small for reliable measurements
(see section \ref{outlook}).
Note that the decrease of $^{63}T_1^{-1}$ should not be
attributed to the presence of a spin-gap.
A clear sign of a (pseudo-) gap would require at least a decrease of 
$(^{63}T_1T)^{-1}$ [$\propto \chi^{\prime\prime}(Q_{AF}$)] with $T$ \cite{NMRreview}.
As seen on Fig. 1(c), this is not the case here. $(^{63}T_1T)^{-1}$
obeyes a Curie, or Curie-Weiss, law.

%%XXX \special{src: 315 MHJPISA3.TEX} %Inserted by TeXtelmExtel

%%XXX \special{src: 318 MHJPISA3.TEX} %Inserted by TeXtelmExtel

%%XXX \special{src: 321 MHJPISA3.TEX} %Inserted by TeXtelmExtel

\subsection{\la~NQR results}
\label{laresults}

%%XXX \special{src: 326 MHJPISA3.TEX} %Inserted by TeXtelmExtel

At low temperatures, \la~NQR can ideally substitute \cu~measurements.
Although La lies outside CuO$_2$ planes, it is coupled to Cu$^{2+}$ spins
through a hyperfine interaction, whose magnitude is small compared
to that on \cu~nuclei. This leads to much longer values of  $^{139}T_1$
and $^{139}T_2$.

%%XXX \special{src: 334 MHJPISA3.TEX} %Inserted by TeXtelmExtel

$^{139}T_1$ was measured by the saturation-recovery method.
Since the form of the time dependence of the \la~magnetization was found
to be $T$-dependent, the $^{139}T_1$ values
reported here are defined as the time at which the magnetization has decreased
by a factor 1/$e$ from its equilibrium value; 
the overall $T$-dependence of $^{139}T_1$ is not affected by the criterion chosen.
By comparing the recovery law of the \la~magnetization after saturation of
the 2$\nu_Q$ transition with that measured on the 3$\nu_Q$ transition, it was found
that the spin-lattice relaxation 
is due to both magnetic {\it and} electric field gradient fluctuations around 100~K.
However,
below $\sim$75~K, 1/$T_1$ increases progressively on cooling and becomes
entirely of magnetic origin.

%%XXX \special{src: 350 MHJPISA3.TEX} %Inserted by TeXtelmExtel

As seen in Fig. 1, $(^{139}T_1)^{-1}$ increases by
almost three orders of magnitude with a peak around $T_g\simeq 5$~K. This behaviour
is typical of a slowing down of spin-fluctuations:
1/$T_1$ reaches a maximum when the frequency of these fluctuations equals the nuclear
resonance frequency, here $\nu_Q\simeq 19$~MHz (or equivalently a correlation time
$\tau \sim 10^{-8}$ s).
A spread of fluctuations frequencies is signaled by the
distribution of $T_1$ values developing below $\sim50$~K (the time decay of
the nuclear magnetization becomes a stretched exponential).

%%XXX \special{src: 362 MHJPISA3.TEX} %Inserted by TeXtelmExtel

In summary, below $\sim 5$~K, in the superconducting state of \lsix,
Cu$^{2+}$ spins are frozen at the NQR time scale of 10$^{-8}$~s, and thus can
be considered as static.
Given the continuous slowing-down characteristic of glass transitions,
a faster probe like neutron scattering should detect an elastic component
\cite{Wakimoto99} appearing at higher temperature.

%%XXX \special{src: 371 MHJPISA3.TEX} %Inserted by TeXtelmExtel

%%XXX \special{src: 374 MHJPISA3.TEX} %Inserted by TeXtelmExtel

%%%%%%%%%%%%%%%%%%%%%%%%%%%%%%%%%%%%%%%%%%%%%%%%%%%%%%%%%%%%%%%

%%XXX \special{src: 378 MHJPISA3.TEX} %Inserted by TeXtelmExtel

\subsection{Comments on the coexistence of the
superconducting and cluster spin-glass phases}
\label{commentcoexistence}

%%XXX \special{src: 384 MHJPISA3.TEX} %Inserted by TeXtelmExtel

When one deals with the coexistence of two phases in the vicinity
of their respective boundaries, it is natural to suspect that the
sample simply contains both phases, well-separated in space.
The most trivial way to obtain two distinct electronic phases
is through a highly non-uniform distribution of Sr dopants.
Such hypothesis can however be ruled out:

%%XXX \special{src: 393 MHJPISA3.TEX} %Inserted by TeXtelmExtel

The first observation is that we find
no sign of a second set of \la~nuclei with longer and 
thermally activated relaxation times (such as usually observed
in the superconducting state).
Thus, we can exclude the presence of two different macroscopic phases
of comparable sizes.
In fact, the vast majority of Cu spins, if not all, becomes frozen.
This does not mean that all \cu~sites have the same $T_1$, but
that their $T$-dependence is similar.
This feature is in agreement with the fact that essentially all
muons experience a static magnetic field below $T_g$
\cite{Kitazawa88,Weidinger89,Niedermayer98}.

%%XXX \special{src: 408 MHJPISA3.TEX} %Inserted by TeXtelmExtel

Furthermore, our crystal has average properties which are well-defined
and are consistent with other measurements:
(i) The freezing temperature $T_g\simeq 5$~K is in agreement with the
NQR \cite{Chou93} and $\mu$SR phase diagrams \cite{Niedermayer98},
carefully established in \lsco~powder samples
(the characteristic times of NQR and $\mu$SR are similar).
(ii) A quantitative analysis of the $T$-dependence of $T_1$ above $T_g$ \cite{Julien99b},
gives values of the characteristic
parameters in agreement with those extracted by Cho \etal~\cite{Cho92}.
(iii) As discussed in \cite{Julien99}, the \cu~$T_1$ values and NQR spectra agree
very well with previously published data.

%%XXX \special{src: 422 MHJPISA3.TEX} %Inserted by TeXtelmExtel

Thus, the magnetic properties of this sample are
characteristic of a single $x=0.06$ phase.
From this and previous experiments
\cite{Niedermayer98,Kitazawa88,Weidinger89,Watanabe94,Kukovitsky95},
it is clear that the spin-freezing
is a robust bulk feature of \lsco~
with $x\leq 0.10$.

%%XXX \special{src: 432 MHJPISA3.TEX} %Inserted by TeXtelmExtel

One may still argue that superconductivity, on the other hand,
is associated with tiny regions having somewhat higher hole concentration.
As to our sample, a clear diamagnetic response is seen in both field cooling
and zero-field cooling magnetization measurements,
below $T_c^{\rm onset}=8$~K, a value in agreement with the 6\% doping level.
However, it is known that no reliable number for the Meissner fraction or the flux
expulsion can be extracted from such measurements in a single crystal and in a field of 
10~Oe (see the discussion in \cite{Nagano93,Tranquada97} for instance).

%%XXX \special{src: 443 MHJPISA3.TEX} %Inserted by TeXtelmExtel

In any event, it is well-known that the superconducting properties of
\lsco~samples with $0.05<x\leq 0.08$ are always rather poor
\cite{Nagano93,Takagi89,Radaelli94,Loram89}.
There is in fact an obvious reason for this:
The superfluid density $\rho_s$ is determined by
$x$ the concentration of doped holes \cite{Uemura89}.
Unavoidable spatial variations of doping
in these non-stochiometric materials thus have a strong influence
at the border of the superconducting phase where $\rho_s$ is small.
A closely related remark is that the coexistence is facilitated
by the fact that it involves, in hole-doped cuprates,
two different electronic entities: localized Cu$^{2+}$ 
spins from which antiferromagnetism arise and doped holes in O$_{2p}$ orbitals
that are the actual actors of superconductivity (in the sense that $\rho_s \propto x$).

%%XXX \special{src: 460 MHJPISA3.TEX} %Inserted by TeXtelmExtel

Even if it remains unclear how homogeneous these superconductors are
(\ie~on which scale the coexistence occurs),
trivial inhomogeneities remain unable to explain 
the occurrence of spin-freezing up to at least $x\simeq 0.10$
\cite{Niedermayer98,Julien99b}, where superconducting
properties are enhanced.
The coexistence has to be intrinsic and to occur on a relatively small lengthscale.
Furthermore as already argued \cite{Weidinger89}, the continuous doping
dependence of both magnetic and superconducting parameters
($T_g$, staggered magnetization, $T_c$) witnesses in favor of a
microscopic phase coexistence.

%%XXX \special{src: 474 MHJPISA3.TEX} %Inserted by TeXtelmExtel

In conclusion, the coexistence of static magnetism with
superconductivity appears to be a reproducible, intrinsic property of
low doped \lsco~($0.05 < x \leq 0.10$).
In this part of the phase diagram, both orders have opposite
doping dependence suggesting, but not proving, that they compete
with each other.

%%XXX \special{src: 483 MHJPISA3.TEX} %Inserted by TeXtelmExtel

%%XXX \special{src: 486 MHJPISA3.TEX} %Inserted by TeXtelmExtel

\subsection{Comment on the \cu~NQR signal}
\label{commentsignal}

%%XXX \special{src: 491 MHJPISA3.TEX} %Inserted by TeXtelmExtel

It may be seen on Fig 1. that 1/$^{63}T_1$ and 1/$^{139}T_1$ data do
not show the same temperature dependence in the range 50~K$<T<100$~K.
The spin-freezing (namely, the increase of 1/$T_1$ with decreasing $T$)
is, unexpectedly, not apparent in $^{63}T_1$ data.
In the light of \la~$T_1$ results, this is in fact not so surprising:
such slow spin fluctuations imply very high values of 1/$T_1$ and 1/$T_2$
on \cu~nuclei, for which the hyperfine coupling to Cu$^{2+}$ spins 
($A$ in equations 1 and 2) is much stronger than for \la~nuclei.
Thus, the recovery towards equilibrium of the magnetization for most
\cu~nuclei is probably too fast to be observed experimentally.
Are only observed those Cu sites that are not yet affected by the slowing-down.
Actually, the disappearance of the \cu~NQR signal for $x\leq\frac{1}{8}$ in \lsco~
was studied very carefully by Hunt \etal~\cite{Hunt99}.
They have shown that the NQR signal disappears because of the combined
effects of short \cu~relaxation times and, more importantly,
of quasi-static charge fluctuations that wipe out the NQR spectrum.
Furthermore, by analogy with similar results in Nd-doped \lsco, Hunt \etal~
could attribute this wipeout effect to charge-stripe ordering \cite{Hunt99}.
Their discovery also explains the loss of NQR/NMR intensity in \lsix,
or at least part of it: It remains to be seen in our sample how much of
the missing signal is due to charge fluctuations, \ie~true wipeout,
or to slow magnetic fluctuations, \ie~too fast nuclear relaxation
(the work of Hunt \etal~suggests that stripe fluctuations vanish in non-superconducting
compositions \cite{Hunt99}).
The missing signal also rationalizes why spin-freezing phenomena have been missed
in some former \cu~NMR/NQR measurements.

%%XXX \special{src: 520 MHJPISA3.TEX} %Inserted by TeXtelmExtel

%%XXX \special{src: 523 MHJPISA3.TEX} %Inserted by TeXtelmExtel

%%XXX \special{src: 526 MHJPISA3.TEX} %Inserted by TeXtelmExtel

\subsection{The charge freezing context}
\label{chargefreezing}

%%XXX \special{src: 531 MHJPISA3.TEX} %Inserted by TeXtelmExtel

It had already been noticed \cite{Harshman88,Hayden91,Julien99,Thurber97}
that the magnetic slowing-down
occurs in the temperature and doping range where in-plane charge transport
shows insulating tendencies, either in zero external magnetic field or under
fields suppressing the effects of the superconducting transition \cite{rem1}
(for $0.04<x<0.08$, the in-plane resistivity $\rho_{ab}$ typically increases below 50-100~K
depending somewhat on disorder \cite{Ando96,Boebinger96,Lai98}).
Here the freezing process is noticeable below $\sim 75$~K, but a precise
assignment of the temperature at which 1/$^{139}T_1$ starts to increase
is not possible since there is just a smooth crossover.
Moreover, 1/$^{139}T_1$ contains a significant background of
quadrupolar relaxation which masks the crossover,
so it may well occur at somewhat higher $T$.
Still, the spin-glass transition $T_g$ and the temperature marking the
minimum of $\rho_{ab}$ have a rather similar doping dependence,
thereby emphasizing the link between spin-freezing and
charge localization
(note that "localization" is used here in the loose sense of a
non-metallic $T$-dependence of $\rho_{ab}$).
In Ref. \cite{Julien99}, a scenario of charge segregation was proposed,
with the motion of the charged domain walls inhibiting the collective
freezing of AF clusters.
Not inconsistent with these views, the results of Hunt \etal~
($T_{\rm charge~order}\simeq 90$~K for La$_{1.93}$Sr$_{0.07}$CuO$_4$ \cite{Hunt99})
now reveal that the insulatinglike state should consist of quasi-static,
still slowly fluctuating, charge objects (for $0.06<x<0.12$).

%%XXX \special{src: 560 MHJPISA3.TEX} %Inserted by TeXtelmExtel

It is interesting to note that there is no detectable enhancement of
the spin-lattice relaxation on approaching the charge ordering.
The amplitude of electric field gradient fluctuations is presumably
too small to produce sizable quadrupolar relaxation.
The absence of a magnetic relaxation peak at the charge ordering
suggests that low-energy magnetic excitations are
not much related to charge motion at this temperatures \cite{oxides}.

%%XXX \special{src: 570 MHJPISA3.TEX} %Inserted by TeXtelmExtel

%%XXX \special{src: 573 MHJPISA3.TEX} %Inserted by TeXtelmExtel

\subsection{Outlook}
\label{outlook}

%%XXX \special{src: 578 MHJPISA3.TEX} %Inserted by TeXtelmExtel

The coexistence of static magnetism (the cluster spin-glass)
with superconductivity can obviously {\it not} be proved from
one set of measurements in a single sample.
Nevertheless, we have enumerated a number of similar and concordant
results, giving credence to the fact that this coexistence is intrinsic
to the low doping side of the superconducting phase.

%%XXX \special{src: 587 MHJPISA3.TEX} %Inserted by TeXtelmExtel

In the bulk of this paper, we have intentionally restricted the
discussion to superconducting compositions {\it below} $x=0.10$.
The reason is that, from $x\simeq 0.05$ up to $x\simeq 0.10$, the most natural reference is 
the cluster spin-glass phase known at lower doping ($0.02 \leq x \leq 0.05$). 
In order to keep a sharp focus, we thus chose not to refer to magnetic freezing
phenomena observed close to the composition $x=0.12$,
in rare-earth doped \lsco, or in La$_{2-x}$Ba$_x$CuO$_4$,
which are more evidently dominated by charge-stripe ordering.
However, the magnetic freezing in, {\it e.g.} \lsix~
has an obvious connection to this physics, which can no longer be ignored
(See Refs. \cite{Julien99,Hunt99,Kivelson98,Singer99}).
The physics of charge segregation provides a natural way 
to understand and unify a number of features, especially
the existence of frozen antiferromagnetic regions at doping levels
as high as those corresponding to superconductivity.
Eventually, the experience with Nd-doped \lsco~with $x\simeq\frac{1}{8}$
makes clear that the coexistence of magnetic and superconducting orders is
possible in hole-doped cuprates \cite{Tranquada97,Ostenson97,Moodenbaugh97,Nachumi98}.

%%XXX \special{src: 608 MHJPISA3.TEX} %Inserted by TeXtelmExtel

Since we tentatively adopted the stripe-glass picture of Emery and Kivelson
\cite{Kivelson98} for $x=0.06$ \cite{Julien99},
one might thus find more appropriate to describe our findings first with respect to
the $x\simeq 0.12$ point, \ie~to well-defined stripes, rather than to the low
doping side where localization and Sr disorder effects might dominate \cite{Gooding97}.
In reality, a lot remains to be done at the experimental level
to better characterize the size and the topology of the magnetic 
and superconducting components across the whole phase diagram \cite{Wakimoto99b},
as well as the associated charge and spin dynamics.
This should obviously help understanding superconductivity at a more microscopic level.

%%XXX \special{src: 621 MHJPISA3.TEX} %Inserted by TeXtelmExtel

%%XXX \special{src: 624 MHJPISA3.TEX} %Inserted by TeXtelmExtel

\subsection{Acknowledgments}
\label{acknowledgments}

%%XXX \special{src: 629 MHJPISA3.TEX} %Inserted by TeXtelmExtel

We are grateful to C.T. Lin (MPI Stuttgart) for providing us with the \lsix~single crystal.
It was a pleasure to discuss these and related topics with A. Rigamonti,
V.J. Emery, R.J. Gooding and Ch. Niedermayer.
The work in Pavia was supported by the INFM-PRA SPIS funding.
Ames Laboratory is operated for U.S Department 
of Energy by Iowa State University under Contract No. W-7405-Eng-82.
The work at Ames Laboratory was supported by the director for
Energy Research, Office of Basic Energy Sciences.

%%XXX \special{src: 640 MHJPISA3.TEX} %Inserted by TeXtelmExtel

%%XXX \special{src: 643 MHJPISA3.TEX} %Inserted by TeXtelmExtel

%%XXX \special{src: 646 MHJPISA3.TEX} %Inserted by TeXtelmExtel

%%XXX \special{src: 1012 MHJPISA3.TEX} %Inserted by TeXtelmExtel


\begin {references}
\bibitem{rem1} Note that an external magnetic field does not 
affect the $T$-dependence of $^{139}T_1$ above $T_g$ in \lsix.
\bibitem{oxides}
This is in contrast to the effects of charge localization in the ordered AF phase 
of other hole-doped transition-metal oxides, like Cu$_{1-x}$Li$_x$O
\cite{Carretta94} and Ni$_{1-x}$Li$_x$O \cite{Corti97}.
However, the charge freezing in \lsco~
is obviously very different from strong charge localization effects 
in AF insulators.
Furthermore, the spin background in \lsco~has
much shorter range correlations (for $T\gg T_g$).
\end{references}
\end{document}